\documentclass[prd,reprint]{revtex4-1}
\usepackage{amsmath,amssymb,amsfonts,amsthm}
\usepackage{mathrsfs}
\usepackage{graphicx}

\begin{document}

\newcommand{\bE}{\bar{E}}
\newcommand{\br}{\bar{r}}
\newcommand{\bt}{\bar{t}}
\newcommand{\btau}{\bar{\tau}}
\newcommand{\blambda}{\bar{\lambda}}
\newcommand{\ba}{\bar{a}}
\newcommand{\bL}{\bar{L}}
\newcommand{\bK}{\bar{K}}
\newcommand{\bp}{\bar{p}}
\newcommand{\bD}{\bar{\Delta}_{\br}}
\newcommand{\bR}{\bar{\mathcal{R}}}
\newcommand{\bT}{\bar{\mathcal{T}}}
\newcommand{\sL}{\textrm{sign}(L)}

\newcommand{\mc}{\mathcal}

\title{Generalized gravitomagnetic clock effect}

\author{Eva Hackmann}
\email{eva.hackmann@zarm.uni-bremen.de}
\affiliation{ZARM, University of Bremen, Am Fallturm, 28359 Bremen, Germany}

\author{Claus L\"ammerzahl}
\email{claus.laemmerzahl@zarm.uni-bremen.de}
\affiliation{ZARM, University of Bremen, Am Fallturm, 28359 Bremen, Germany}
\affiliation{Institute for Physics, University Oldenburg, 26129 Oldenburg, Germany}

\begin{abstract}
In General Relativity, the rotation of a gravitating body like the Earth influences the motion of orbiting test particles or satellites in a non-Newtonian way. This causes, e.g., a precession of the orbital plane known as the Lense-Thirring effect and a precession of the spin of a gyroscope known as the Schiff effect. Here, we discuss a third effect first introduced by Cohen and Mashhoon called the gravitomagnetic clock effect. It describes the difference in proper time of counterrevolving clocks after a revolution of $2\pi$. For two clocks on counterrotating equatorial circular orbits around the Earth, the effect is about $10^{-7}$ seconds per revolution, which is quite large. We introduce a general relativistic definition of the gravitomagnetic clock effect which is valid for arbitrary pairs of orbits. This includes rotations in the same direction and different initial conditions, which are crucial if the effect can be detected with existing satellites or with payloads on nondedicated missions. We also derive the post-Newtonian expansion of the general relativistic expression and calculate the effect for the example of a satellite of a Global Navigation Satellite System compared to a geostationary satellite. 
\end{abstract}

\maketitle

\section{Introduction}
Within Einstein's General Relativity the rotation of an astronomical object like the Earth has a purely relativistic effect on the motion of orbiting objects which is usually referred to as ``frame dragging''. Maybe better terminology can be found in analogy to electrodynamics by denoting the effects due to mass currents as ``gravitomagnetic''. Both terms summarize at least two well-known effects. First, the Lense-Thirring effect calculated in 1918 \cite{LenseThirring18,Mashhoonetal_LT}, which in the weak field can be interpreted as a precession of the longitude of the ascending node. Solar System measurements of the Lense-Thirring precession were achieved with the LAGEOS mission and will be further improved using the LARES satellite, see e.g.~\cite{Ciufolinietal2013}. Second, the Schiff effect calculated in 1960 \cite{Schiff1960,Schiff1960b}, which describes the precession of a gyroscope orbiting a rotating object. This effect was measured by the Gravity Probe B experiment \cite{Pugh1959,Everittetal2011}.

In this paper we are interested in another effect caused by the rotation of the gravitating object, the so-called gravitomagnetic clock effect. It refers to different time measurements of two clocks orbiting a rotating astronomical object, one in the direction of rotation, i.e.~on a prograde orbit, and the other against the direction of rotation, i.e.~on a retrograde orbit. There are several versions of this clock effect which differ in the details of their definitions \cite{Binietal2001,Mashhoonetal1999,Tartaglia2000}. Here, we discuss what the authors of \cite{Binietal2001} call the observer-dependent two-clock clock effect. This is the difference of the proper times of two clocks, one on a pro- and the other on a retrograde orbit, after a complete revolution of $2\pi$. Note that the proper time measured by a clock on a geodesic is invariant. The observer enters the discussion through the notion of a ``full revolution'' which depends on the frame of reference \cite{BonnorSteadman1999}. Alternative definitions of the gravitomagnetic clock effect include e.g. the difference in the proper time of the two clocks after a fixed coordinate time of an observer \cite{Mashhoonetal1999} or the difference in proper time of the two clocks at their meeting point \cite{Tartaglia2000}.

The gravitomagnetic clock effect considered here was first correctly derived and studied in detail by Cohen and Mashhoon \cite{CohenMashhoon1993} following an idea shortly mentioned in \cite{Rosenblum1987,Cohenetal1988}. For two counterrevolving clocks on circular orbits of the same radius in the equatorial plane of the Earth Cohen and Mashhoon found that the effect is of the order of $10^{-7}$ seconds per revolution, independent of the radius of the two clocks. Compared to the increasing accuracies of space-based clocks this seems to be quite large. A generalization to the parametrized post-Newtonian formalism also including the effects of the nonspherical shape of the Earth was considered in \cite{Mashhoonetal2001b}. Eccentric and inclined orbits were discussed in \cite{Mashhoonetal1999, Mashhoonetal2001}, but the requirement of identical initial orbital parameters, apart from the sense of rotation, was so far not removed. A dedicated satellite mission to measure this effect called Gravity Probe $\rm C_{lock}$ was proposed by Gronwald and others \cite{Gronwaldetal1997}. Gravitational and nongravitational error sources for such a mission were also discussed \cite{Gronwaldetal1997,Iorio2001,Iorio2001b,Iorioetal2005,Lichteneggeretal2006}. From these analyses, it can be concluded that the most challenging task for a mission to measure the clock effect is not the stability of the orbiting clocks but the precise tracking of the two satellites. This is needed because of the imperfect cancellation of the Keplerian periods of the two clocks, which induces large errors in the measurement. 

In this paper, we find a fully general relativistic definition of the gravitomagnetic clock effect in Kerr spacetime, which also generalizes the clock effect to two arbitrary geodesics, including rotations in the same direction and different initial orbital parameters (see also \cite{Hackmannetal2013}). We use for this definition the fundamental frequencies of bound orbits in Kerr spacetime given by Schmidt \cite{Schmidt02} and elaborated by Fujita and Hikida \cite{FujitaHikida09}. This procedure is completely analogous to the definition of the perihelion shift and the Lense-Thirring effect in terms of fundamental frequencies \cite{FujitaHikida09,HackmannLaemmerzahl2012}. The generalization of the gravitomagnetic clock effect to two arbitrary geodesics would, in principle, allow to use existing satellites for a measurement of the effect as long as they carry stable clocks and can be tracked with sufficient accuracy. We also derive a post-Newtonian expansion of the generalized gravitomagnetic clock effect which can be handled more conveniently than the fully general relativistic expression and should still be sufficiently accurate for orbits in the gravitational field of the Earth. This expression is then applied to a spacecraft of a Global Navigation Satellite System (GNSS) compared to a geostationary satellite. The paper closes with a summary.

\section{Fundamental frequencies in Kerr spacetime}\label{sec:frequencies}
We start with a review and an extension of the fundamental frequencies in Kerr spacetime given by Schmidt \cite{Schmidt02} and by Fujita and Hikida \cite{FujitaHikida09}. The Kerr metric in Boyer-Lindquist coordinates reads
\begin{align}\label{K_metric}
ds^2 & = \frac{\Delta_r}{\rho^2} \left(cdt - a \sin^2\theta d\varphi\right)^2 - \frac{\rho^2}{\Delta_r} dr^2  - \rho^2 d\theta^2\nonumber\\
& \quad - \frac{\sin^2\theta }{\rho^2} (a cdt - (r^2+a^2) d\varphi)^2\,,
\end{align}
where $ds^2=c^2d\tau^2$ with the proper time $\tau$, $\Delta_r = r^2+a^2-2M r$, $\rho^2 = r^2+a^2\cos^2\theta$ with $M=Gm/c^2$ and $a=J/(mc)$, where $m$ is the mass and $J>0$ is the angular momentum of the gravitating object. Here $G$ is the gravitational constant and $c$ the speed of light.

The geodesic equation in Kerr spacetime can be completely separated due to the existence of four constants of motion. This is the specific energy $\tilde{E}$ and the specific angular momentum in the direction of the symmetry axes $\tilde{L}_z$,
\begin{align}
\tilde{E} & = g_{tt} \dot t + g_{t\varphi} \dot \varphi =: c^2E\,,\\
\tilde{L}_z & = - g_{\varphi \varphi} \dot \varphi - g_{t\varphi} \dot t =: cL_z\,,
\end{align}
where the dot denotes a derivative with respect to $\tau$. The two remaining constants are the normalization $g_{\mu\nu}\dot{x}^\mu\dot{x}^\nu=\epsilon c^2$ with $\epsilon=1$ for massive test particles and the Carter constant $K$ \cite{Carter68}. There are some alternative forms of the Carter constant; we use here $K$ such that $K=(aE-L_z)^2$ for motion in the equatorial plane. 

With these constants of motion we get the equations of motion in the form \cite{Mino03}
\begin{align}
\left( \frac{dr}{d\lambda} \right)^2 & = \mathcal{R}^2 - \Delta_r (\epsilon r^2 + K) =: R\,, \label{dot r_lambda} \\
\left( \frac{d \theta}{d\lambda} \right)^2 & = K- \epsilon a^2 \cos^2\theta - \frac{\mathcal{T}^2}{\sin^2\theta} =: \Theta\,, \label{dot theta_lambda} \\
\frac{d \varphi}{d \lambda} & = \frac{a}{\Delta_r} \mathcal{R} - \frac{\mathcal{T}}{\sin^2\theta}=:\Phi_r(r)+\Phi_\theta(\theta)\,, \label{dot phi_lambda} \\
c\frac{d t}{d\lambda} & = \frac{r^2+a^2}{\Delta_r} \mathcal{R}  - a \mathcal{T} =: T_r(r)+T_\theta(\theta)\,. \label{dot t_lambda}
\end{align}
where
\begin{align}
\mathcal{R} & = (r^2+a^2) E - a L_z\,,\\
\mathcal{T} & = aE\sin^2\theta-L_z\,.
\end{align}
Here $\lambda$ is the ``Mino time'' which is connected to the proper time by $cd\tau = \rho^2 d\lambda$. It is an auxiliary parameter introduced to completely decouple the equations of motions. Note that the equations of motions can be rewritten in a dimensionless form by dividing each by the appropriate power of $M$ and redefining
\begin{align}
\br & =r/M\,,\quad \bar t=ct/M\,, \quad \ba=a/M\,,\nonumber\\
\bL_z & =L_z/M\,,\quad \bK=K/M^2\,,\quad \bar\lambda=\lambda M\,.
\end{align}

In general, the motion of test particles in Kerr spacetime neither has the form of a conic section nor lies in an orbital plane. This is due to a mismatch of the periodicities of the radial and latitudinal motion, which, in general, differ from each other and from $2\pi$ \cite{Schmidt02,FujitaHikida09}. Let us consider a bound orbit of a massive test particle (i.e.~$\epsilon=1$) which does not cross a horizon. In this case the radial motion oscillates between the periapsis $r_{\rm p}$ and the apoapsis $r_{\rm a}$. Also the test particle oscillates around the equatorial plane between two extremal values $\theta_{\rm min,max}$ with $\theta_{\rm max} = \pi - \theta_{\rm min}$. The radial and latitudinal periods $\varLambda_{r,\theta}$ are then given by
\begin{align}
\varLambda_r & = 2 \int_{\br_{\rm p}}^{\br_{\rm p}} \frac{d\br}{\sqrt{R(\br)}}\,,\\
\varLambda_\theta & = 2\int_{\theta_{\rm min}}^{\theta_{\rm max}} \frac{d\theta}{\sqrt{\Theta(\theta)}}\,,
\end{align}
i.e.~$\br(\blambda+\varLambda_r)=\br(\blambda)$ and $\theta(\blambda+\varLambda_\theta)=\theta(\blambda)$ for all $\blambda$. The conjugate fundamental frequencies are defined as $\Upsilon_r=2\pi/\varLambda_r$ and $\Upsilon_\theta=2\pi/\varLambda_\theta$.

As the $\varphi$, $t$, and $\tau$ motions are not periodic we have to use a somewhat different approach to define the corresponding fundamental frequencies. We write the coordinate as a part which is linear in $\lambda$ plus perturbations in $r$ and $\theta$,
\begin{align}
\varphi(\blambda) & = \Upsilon_\varphi \blambda + \Phi^r_{\textrm osc}(\br) + \Phi^\theta_{\textrm osc}(\theta)\,,\\
\Upsilon_\varphi & := \langle \Phi_r(\br)+\Phi_\theta(\theta) \rangle_{\blambda}\,, \label{defYphi}
\end{align}
where 
\begin{equation}
\langle \cdot \rangle_{\blambda}:=\lim_{(\blambda_2-\blambda_1) \to \infty} \frac{1}{\blambda_2-\blambda_1} \int_{\blambda_1}^{\blambda_2} \cdot \, d\blambda \label{avlambda}
\end{equation} 
is an infinite time average with respect to $\blambda$. The functions $\Phi^r_{\textrm osc}(\br)$ and $\Phi^\theta_{\textrm osc}(\theta)$ represent oscillatory deviations from the average. They are defined by
\begin{align}
\Phi^r_{\textrm osc}(\br) & = \int \Phi_r(\br)  d\blambda - \left\langle \Phi_r(\br) \right\rangle_{\blambda} \blambda \,, \\
\Phi^\theta_{\textrm osc}(\theta) & = \int \Phi_\theta(\theta) d\blambda - \left\langle \Phi_\theta(\theta) \right\rangle_{\blambda} \blambda
\end{align}
and have periods $\varLambda_r$ and $\varLambda_\theta$, respectively.

As $\Upsilon_\varphi$ contains $\br$- and $\theta$-dependent parts which are periodic functions with respect to $\blambda$, the integral \eqref{avlambda} can be reduced to an integral over one period. Therefore we find
\begin{align}
\Upsilon_\varphi & = \frac{2}{\varLambda_r} \int_{\br_{\rm p}}^{\br_{\rm a}} \frac{\Phi_r(\br) d\br}{\sqrt{R(\br)}} + \frac{2}{\varLambda_\theta} \int_{\theta_{\rm min}}^{\theta_{\rm{max}}} \frac{\Phi_\theta(\theta) d\theta}{\sqrt{\Theta(\theta)}}\,. \label{Upsilonphi}
\end{align}
Analogously, we may define
\begin{align}
\Upsilon_t & = \frac{2}{\varLambda_r} \int_{\br_{\rm p}}^{\br_{\rm a}} \frac{T_r(\br) d\br}{\sqrt{R(\br)}} + \frac{2}{\varLambda_\theta} \int_{\theta_{\rm min}}^{\theta_{\rm{max}}} \frac{T_\theta(\theta) d\theta}{\sqrt{\Theta(\theta)}}\,,\label{Upsilont}\\ 
\Upsilon_\tau & = \frac{2}{\varLambda_r} \int_{\br_{\rm p}}^{\br_{\rm a}} \frac{\br^2 d\br}{\sqrt{R(\br)}} + \frac{2}{\varLambda_\theta} \int_{\theta_{\rm min}}^{\theta_{\rm{max}}} \frac{a^2\cos^2\theta d\theta}{\sqrt{\Theta(\theta)}}\,.\label{Upsilontau}
\end{align}

\section{General definition of the gravitomagnetic clock effect}
The gravitomagnetic clock effect considered here was studied already in 1993 by Cohen and Mashhoon \cite{CohenMashhoon1993}. They showed that two clocks on circular equatorial orbits of the same radius but orbiting in different directions show after a revolution of $2\pi$ a time difference of 
\begin{equation}
\tau_+ - \tau_- \approx 4 \pi \frac{J}{m c^2}\,, \label{deltatauPN}
\end{equation}
where $\tau_{+}$ is the proper time of the corotating and $\tau_-$ of the counterrotating clock. Here $J$ is the angular momentum of the Kerr black hole, as before. For satellites orbiting the Earth this yields an effect of the order of $10^{-7}\,{\rm s}$ per revolution, which is surprisingly large. The key element is here the measurement after a full revolution of $2\pi$. For measurements after a specific coordinate time or at the meeting point of the clocks the effect is much smaller \cite{Mashhoonetal1999,Tartaglia2000}. The result \eqref{deltatauPN} was generalized to spherical orbits with small inclination in \cite{Mashhoonetal1999} and further to orbits with small eccentricity in \cite{Mashhoonetal2001} (their equation (31)),
\begin{align}
t_+ - t_- & \approx 4\pi\frac{J \cos i}{mc^2} \bigg[\frac{-3}{\sqrt{1-e^2}} + \nonumber\\
& \qquad + \frac{4-2\cos^2\varphi_0\tan^2i}{(1+e\cos(\varphi_0-g))^2} \bigg]\,, \label{deltatauPNgen}
\end{align}
where $i$ is the inclination measured from the equatorial plane, $e$ is the eccentricity, $g$ is the argument of the pericenter, and $\varphi_0-g$ is the true anomaly at $t=0$. As remarked in \cite{Mashhoonetal2001}, at the considered level of approximation the coordinate times used in \eqref{deltatauPNgen} may be replaced by the proper times. Note that the expression \eqref{deltatauPNgen} depends on the initial position of the two clocks.

Here we introduce a fully general relativistic expression for the gravitomagnetic clock effect based on fundamental frequencies. As outlined in section \ref{sec:frequencies} the functions $\varphi(\blambda)$, $\bt(\blambda)$, and $\btau(\blambda)$ can be written as a part which is linear in $\blambda$ plus periodic perturbations. As only the linear parts contribute to the average secular increase of the coordinate, we may use this to define observable quantities like the perihelion shift and the Lense-Thirring effect \cite{FujitaHikida09,HackmannLaemmerzahl2012}. Analogously, for the gravitomagnetic clock effect we may define a function $\btau: \varphi \mapsto \btau(\blambda(\varphi))$ by using the linearized functions $\blambda(\varphi)=\Upsilon_\varphi^{-1} \varphi$, $\btau(\blambda)=\Upsilon_\tau \blambda$,
\begin{align}
\btau(\varphi) := \Upsilon_\tau \Upsilon_\varphi^{-1} \varphi\,.\label{deftauphi}
\end{align}
In the Newtonian limit, $\tau(2\pi)$ (as well as the corresponding $t(2\pi)$) reduces to the Keplerian time of revolution $2\pi\sqrt{\frac{d^3}{Gm}}$, where $d$ is the semimajor axis; see \eqref{taugm} below. 

Assume now two clocks moving along arbitrary geodesics with given periapsides $r_{{\rm p},n}$, apoapsides $r_{{\rm a},n}$, and maximal inclinations $\theta_{{\rm max},n}$, $n=1,2$. For each orbit we may calculate the proper time for a revolution of $\pm 2\pi$ using \eqref{deftauphi}, $\btau_{n}(\pm2\pi;a)$, where the sign in front of $2\pi$ indicates pro- ($+$) or retrograde ($-$) motion and the additional argument indicates the dependence of $\btau$ on the Kerr rotation parameter $a>0$. We may also calculate the corresponding value in case the rotation of the central object would vanish, $\btau_{n}(\pm2\pi;0)$. To extract the purely \textit{gravitomagnetic} effect, we define a new observable
\begin{align}
\Delta\btau_{\rm gm} := \btau_{1}(\pm2\pi)+\alpha \btau_{2}(\pm2\pi)\,, \label{deltatau}
\end{align}
where the factor of proportionality $\alpha$ is calculated such that the usual \textit{gravitoelectric} effects just cancel each other. This condition determines $\alpha$ via $\Delta\btau_{\rm gm}=0$ for $a=0$,
\begin{align}
0= \btau_1(\pm2\pi;0) + \alpha \btau_2(\pm2\pi;0)
\end{align}
and, therefore,
\begin{align}
\alpha = - \frac{\btau_1(\pm2\pi;0)}{\btau_2(\pm2\pi;0)}\,. \label{defalpha}
\end{align}
The sign in front of $2\pi$ has to be chosen for each orbit according to its sense of rotation, i.e.~$+2\pi$ ($-2\pi$) for prograde (retrograde) orbits.

The actual calculation procedure for $\Delta\tau_{\rm gm}$ is then as follows: for both clocks, calculate the energies $E_n$, the angular momenta $L_{z,n}$, and the Carter constants $K_n$, $n=1,2$, by using $\frac{dr}{d\lambda}(r_{\rm p,a})=0$ in \eqref{dot r_lambda} and $\frac{d\theta}{d\lambda}(\theta_{\rm max})=0$ in \eqref{dot theta_lambda}. As one may choose $E>0$ without loss of generality, for each orbit this gives two solutions, one for a prograde orbit with $L_z>0$ and one for a retrograde orbit with $L_z<0$, from which we choose according to their sense of rotation. Note that the values $E_n$, $L_{z,n}$ and $K_n$ depend on the rotation parameter $a$ of the central object, $E_n=E_n(a)$, $L_{z,n}=L_{z,n}(a)$, $K_n=K_n(a)$. The corresponding values $E_n(0)$, $L_{z,n}(0)$, $K_n(0)$ can be determined by setting $a=0$ in \eqref{dot r_lambda} and \eqref{dot theta_lambda}. Then use \eqref{Upsilonphi} and \eqref{Upsilontau} (or the corresponding expressions in terms of Jacobian elliptic integrals given in appendix \ref{app:FF}) to calculate $\btau_n(\pm2\pi;a)$ and $\btau_n(\pm2\pi;0)$, which gives $\alpha$ and $\Delta\btau_{\rm gm}$.

Note that the value of the clock effect \eqref{deltatau} depends on the numeration of the two clocks. If we denote by $\Delta\btau_{\rm gm}^{(2,1)}$ the clock effect with reversed clock labels as compared to $\Delta\btau_{\rm gm}^{(1,2)}$ we find
\begin{align}
\Delta\btau_{\rm gm}^{(2,1)} = - \frac{\btau_2(\pm2\pi;0)}{\btau_1(\pm2\pi;0)} \Delta\btau_{\rm gm}^{(1,2)}\,.
\end{align}
This ambiguity can be removed if $\Delta\btau_{\rm gm}$ is referred to the Schwarzschild orbit time of the first clock,
\begin{align}
\frac{\Delta\btau_{\rm gm}}{\btau_1(\pm2\pi;0)} & = \frac{\btau_1(\pm2\pi;a)}{\btau_1(\pm2\pi;0)} - \frac{\btau_2(\pm2\pi;a)}{\btau_2(\pm2\pi;0)}\,.
\end{align}
The absolute value of this quantity does not change if the two clocks interchange their labels.

\section{Post-Newtonian expansion}

We explore now the definition \eqref{deltatau} by deriving an expansion, where we assume that the rotation parameter $a$ and the mass parameter $M$ are small compared to the radii of the clock orbits. This holds for the exterior gravitational field of the Earth where $M/r \lesssim 7 \times 10^{-10}$ and $a/r \lesssim 6 \times 10^{-7}$.

Let us assume that the orbital parameters $r_{\rm p}$, $r_{\rm a}$ and $\theta_{\rm max}$ are fixed for both orbits. Therefore, to derive the expansion of $\tau(\pm2\pi)$ for small $a$ we need the expansions of the constants of motions in terms of $a$, which are given with some additional details in appendix \ref{app:PN}. We find
\begin{align}
\tau(\pm2\pi) & \approx 2\pi\sqrt{\frac{d^3}{Gm}} \left( 1 - \frac{3(1+e^2)}{2(1-e^2)} \frac{M}{d} \right) \label{taugm} \\
& \quad \pm \frac{2\pi(\cos i (3e^2+2e+3) -2e-2)}{(1-e^2)^\frac32} \frac{a}{c}\,,\nonumber
\end{align}
where $d$ is the semimajor axis, $e$ is the eccentricity, and $i$ is the inclination, which are defined via $r_{\rm p}=d(1-e)$, $r_{\rm a}=d(1+e)$, and $\theta_{\rm max}=\pi/2+i$. The sign in \eqref{taugm} has to be chosen according to the sense of rotation, the plus (minus) sign for prograde (retrograde) motion. 

\begin{figure}
\includegraphics[width=0.45\textwidth]{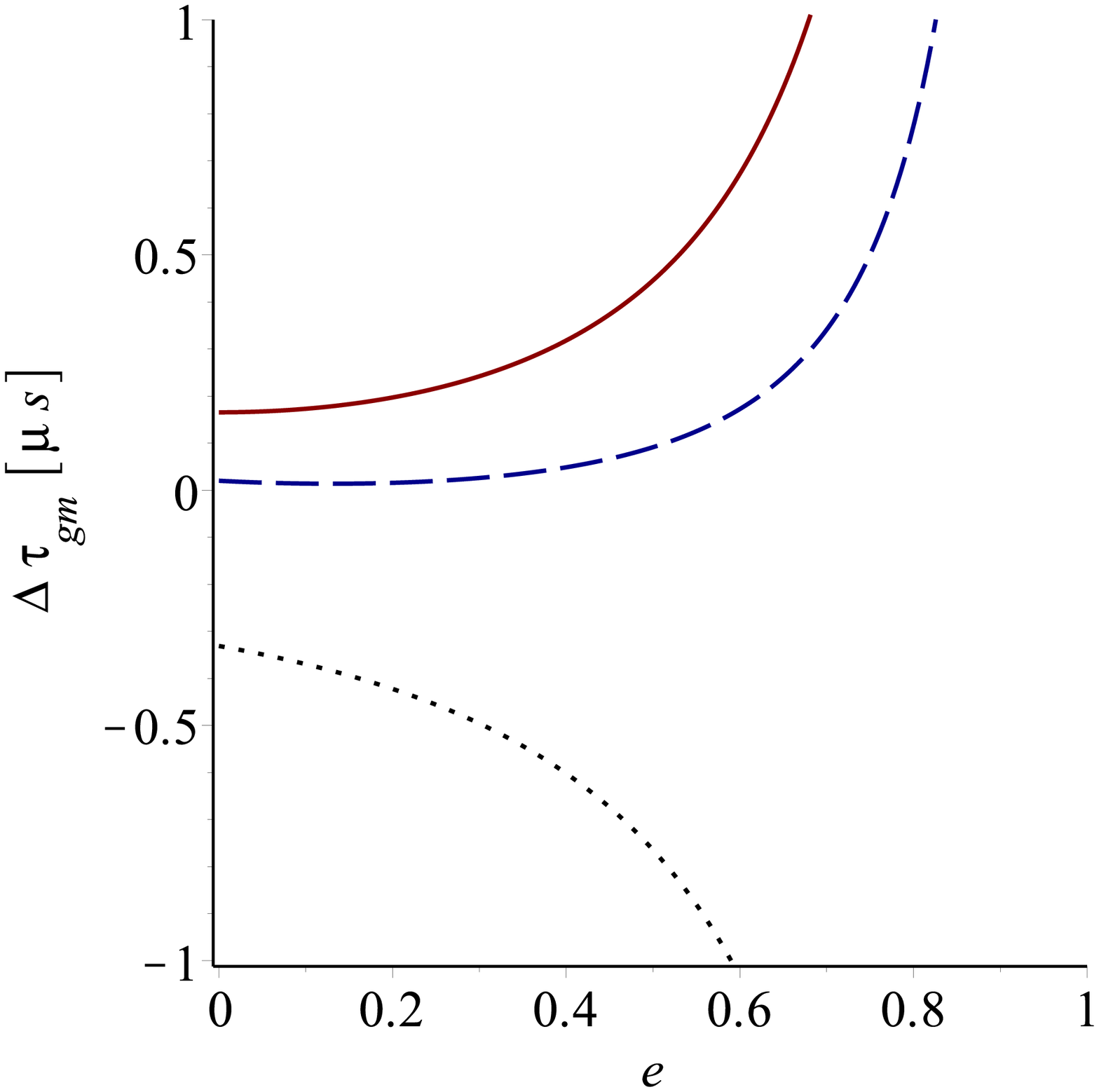}
\includegraphics[width=0.45\textwidth]{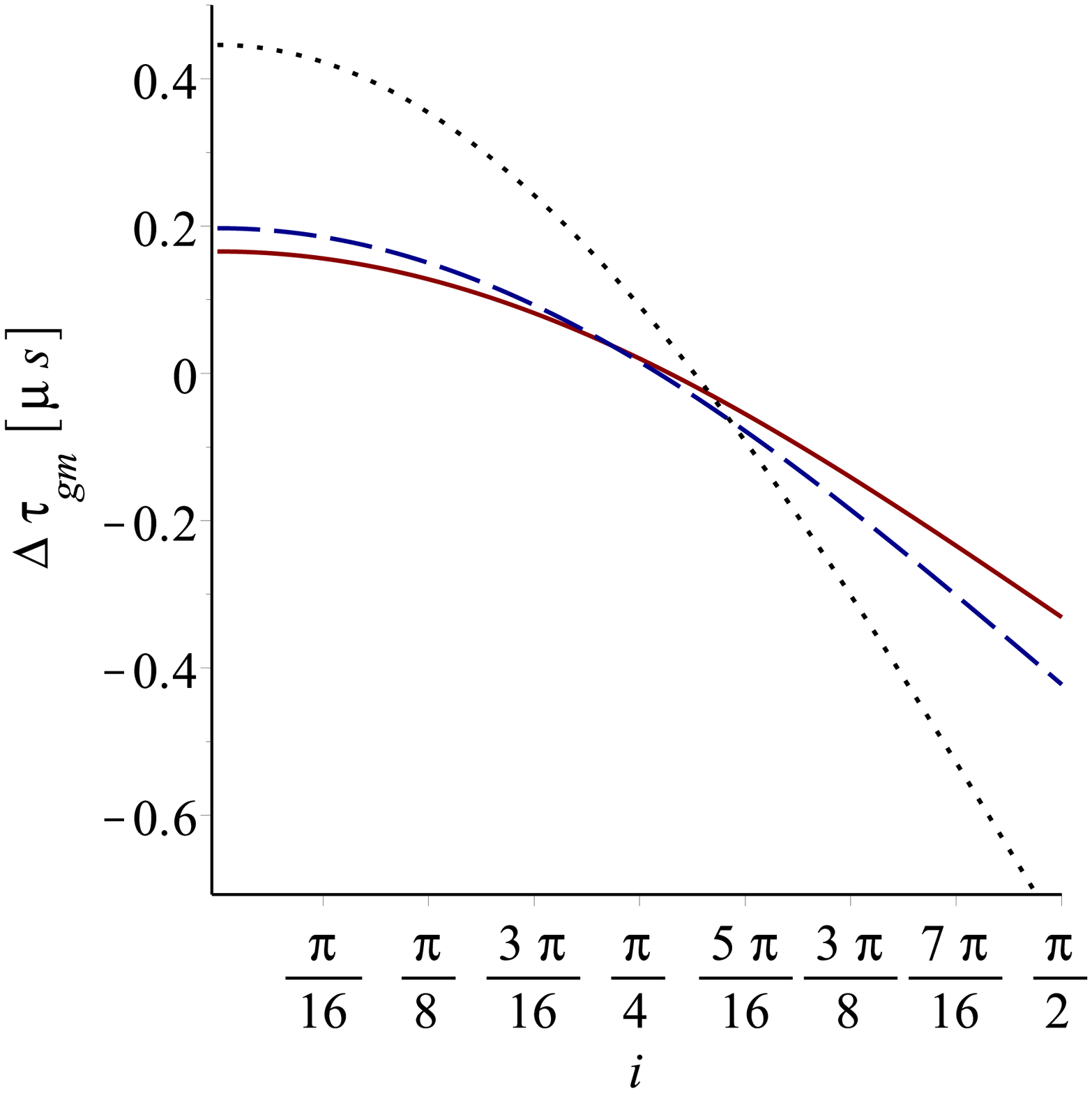}
\caption{The gravitomagnetic clock effect \eqref{PNgm} around the Earth. Top: The solid line is for equatorial orbits, the dashed line for inclination $i=\pi/4$, and the dotted line indicates the limit values for $i\to\pi/2$. Bottom: The solid line is for spherical orbits, the dashed line for eccentricity $e=0.2$, and the dotted line for $e=0.5$.}
\label{Fig:PNexp}
\end{figure}

Consider now the gravitomagnetic clock effect for two clocks on geodesics with identical orbital parameters $r_{\rm p}$, $r_{\rm a}$, and $\theta_{\rm max}$, one on a pro- and the other on a retrograde orbit. Here $\alpha=-1$ and with $\tau_+=\tau(+2\pi)$ and $\tau_-=\tau(-2\pi)$ we find
\begin{align}
\tau_+-\tau_- & \approx \frac{4\pi J}{mc^2} \, \frac{\cos i (3e^2+2e+3) -2e-2)}{(1-e^2)^\frac32}\,,\label{PNgm}
\end{align}
which for circular equatorial geodesics ($e=0$, $i=0$) reduces to the formula \eqref{deltatauPN} derived in \cite{CohenMashhoon1993}. For fixed values of the inclination the effect is visualized in figure \ref{Fig:PNexp}. At a fixed eccentricity the clock effect \eqref{PNgm} vanishes if
\begin{align}
\cos i = \frac{2(1+e)}{3e^2+2e+3}
\end{align}
and gets negative for larger inclinations, see also figure \ref{Fig:PNexp}. Note that the expansion \eqref{taugm} is not valid for polar orbits, see appendix \ref{app:PN}. However, for nearly polar orbits the expression \eqref{taugm} holds and the clock effect \eqref{PNgm} approaches a maximal absolute value,
\begin{align}
\tau_--\tau_+ & \to \frac{8\pi J}{mc^2} \, \frac{1+e}{(1-e^2)^\frac32}\,.
\end{align}

Equation \eqref{PNgm} differs from the expression \eqref{deltatauPNgen} in a key aspect: it does not depend on the initial position of the clocks. This is due to the definition via fundamental frequencies, which are given by averages over infinite Mino time. This procedure is completely analogous to the derivation of the Lense-Thirring effect $\Omega_{\rm LT}$ in the fully general relativistic setting \cite{FujitaHikida09, HackmannLaemmerzahl2012},
\begin{align}
\Omega_{\rm LT} & = \Upsilon_\varphi \varLambda_\theta - 2\pi
\end{align}
which in the post-Newtonian expansion (see appendix \ref{app:PN} for details) reduces to $\Omega_{\rm LT} \approx 4\pi/(d(1-e^2))^\frac32) aM^\frac12$ which is the classical result \cite{LenseThirring18} if it is referred to the Newtonian orbit time $2\pi \sqrt{d^3/(Gm)}$.

Also, \eqref{PNgm} is valid for any $0\leq e<1$ and $0\leq i<\pi/2$, whereas \eqref{deltatauPNgen} assumes small eccentricities and inclinations. But also in the limit of small $e$ or small $i$ the expressions \eqref{PNgm} and \eqref{deltatauPNgen} differ. This is, of course, natural as \eqref{PNgm} is independent of the initial conditions.

For two general orbits we derive a post-Newtonian expression of $\alpha$ (see \eqref{defalpha}),
\begin{align}
\alpha & \approx - \frac{d_1^\frac32}{d_2^\frac32} - \frac{3d_1^\frac12}{2d_2^\frac52} \left[ \frac{d_1(1+e_2^2)}{1-e_2^2}-\frac{d_2(1+e_1^2)}{1-e_1^2} \right] M \label{PNalpha}
\end{align}
and from \eqref{taugm} the post-Newtonian expansion of the gravitomagnetic clock effect to lowest order,
\begin{align}
\Delta\tau_{\rm gm} & \approx \frac{a}{c} \bigg[ s_1 \frac{2\pi(\cos i_1 (3e_1^2+2e_1+3) -2e_1-2)}{(1-e_1^2)^\frac32}\nonumber\\
& \, -s_2 \frac{2\pi d_1^\frac32(\cos i_2 (3e_2^2+2e_2+3) -2e_2-2)}{d_2^\frac32(1-e_2^2)^\frac32} \bigg]\label{genPNgm}
\end{align}
where the indices indicate the first/second clock. Here $s_1$ and $s_2$ are equal to $+1$ (prograde motion) or $-1$ (retrograde motion) according to the sense of rotation of the respective orbit. Note that the expressions \eqref{genPNgm} and \eqref{taugm} diverge for $e\to1$, which is not surprising as $r_{\rm p}=d(1-e)>r_B$, where $r_B$ is the radius of the central body, requires $d\to\infty$ for $e\to 1$. Therefore, in this limit the orbit time itself diverges and it does not make sense to consider the clock effect.

\section{Application to satellites of the GNSS}

We apply now the post-Newtonian expression \eqref{genPNgm} for the calculation the gravitomagnetic clock effect to satellites of the Global Navigation Satellite Systems (GNSS), which carry very stable clocks with frequency stabilities of about $10^{-14}$ over ten thousand seconds. For a detection of the effect, the proper times of the satellite clocks after a revolution of $2\pi$ have to be communicated to an observer on the ground, who also determines the orbital parameters $d$, $e$, and $i$ for each clock. Note that the expression \eqref{taugm} relates these parameters and, therefore, may be used for a consistency check. Using $d$ and $e$ of both clocks we then calculate $\alpha$ by \eqref{PNalpha} and then the gravitomagnetic clock effect \eqref{deltatau}.

As all GNSS satellites are on nearly circular prograde orbits we need a pair of satellites with different inclinations. The GPS and Galileo satellite systems operate on very similar inclination ($55^\circ$ and $56^\circ$, respectively) whereas the GLONASS system operates at slightly larger inclination ($65^\circ$), which is still quite close to the GPS and Galileo systems. The Chinese COMPASS system, however, includes geostationary satellites. Therefore, we compare here a geostationary orbit with a Galileo and a GLONASS satellite orbit.

For the Galileo satellites we assume an inclination of $i_{\rm Ga}=56^\circ$, an eccentricity of $e=0$, and a semimajor axis of $d_{\rm Ga} = 29593 \, \rm km$, and for the GLONASS system $i_{\rm GL}=64.8^\circ$, $e=0$, and $d_{\rm GL}=25471\, \rm km$. As orbital parameters of the geostationary satellite we take $i_{\rm Ge}=0$, $e=0$, and $d_{\rm Ge}=42157 \, \rm km$.

We insert this into the expression \eqref{genPNgm} together with the mass and rotation parameters of the Earth, $M\approx 4.4346 \times 10^{-3} \,\rm m $ and $\frac{a}{c} \approx 1.317 \times 10^{-8} \, \rm sec$. For the Galileo satellite (clock 1) compared to the geostationary satellite (clock 2) we find
\begin{align}
\Delta\tau_{\rm gm} \approx -7.54 \times 10^{-8} \, \rm sec\,.
\end{align}
Referred to the Schwarzschild orbit time of the Galileo satellite we get
\begin{align}
\frac{\Delta\tau_{\rm gm}}{\tau_{\rm Ga}(\pm2\pi;0)} & \approx -1.49 \times 10^{-12}\,.
\end{align}
For the GLONASS satellite we find
\begin{align}
\Delta\tau_{\rm gm} & \approx -9.87 \times 10^{-8} \, \rm sec\,,\\
\frac{\Delta\tau_{\rm gm}}{\tau_{\rm GL}(\pm2\pi;0)} & \approx -2.44 \times 10^{-12}\,.
\end{align}
In principle, the clocks on board the GNSS satellites, including the geostationary COMPASS satellite, should be able to detect this effect. However, we assumed here geodesic motion in a Kerr spacetime in the absence of any disturbing forces. Therefore, a careful analysis of the influence of gravitational and environmental perturbations has to be carried through to judge the measurability of the effect. An analysis of disturbing effects for the situation of identical but counterrevolving clock orbits can be found in \cite{Gronwaldetal1997,Iorio2001,Iorio2001b,Iorioetal2005,Lichteneggeretal2006}. Let us only note two major points here. First, the gravitomagnetic clock effect is quite large compared to the sensitivity of the clocks but tiny compared to the measured proper times for a full revolution, see also \cite{Lichteneggeretal2006}. If we assume an uncertainty in the semimajor axis of $\Delta d$, then $\Delta\tau/\tau \approx 3/2 \, \Delta d/d$, which implies that in the above examples the semimajor axes must be known with an accuracy of about $10 \, \rm \mu m$. As the gravitomagnetic clock effect accumulates over every revolution this stringent requirement can be relaxed with sufficiently long observation times. Another theoretical possibility to achieve a high accuracy for the value of $d$ would be to use a second clock in each satellite, whose position with respect to the first clock is very well known. If the measured proper time of this second clock, say $\tau'(\pm2\pi)$, is inserted in equation \eqref{taugm} we may calculate its semimajor axis $d'$, assuming $e'$ and $i'$ for this clock are known to sufficient precision. If $d'=d+\delta d$, where $\delta d$ is very well known, we may calculate the semimajor axis $d$ of the first clock which is used to detect the gravitomagnetic effect. Second, the high accuracy for the inclination mentioned in \cite{Iorioetal2005} does not apply for the situation considered here, as we assumed different inclinations for the two satellites. If we assume an uncertainty of $\Delta(\cos i)$, we find for orbits of eccentricity $e=0$ that $\Delta\tau \approx 6\pi \frac{a}{c} \Delta(\cos i)$, which implies that the inclinations in the examples above should be known to an accuracy of at least $0.03$ degrees.

\section{Summary}

We presented a generalization of the gravitomagnetic clock effect \cite{CohenMashhoon1993} for two clocks moving on arbitrary geodesics in the Kerr spacetime. The definition uses the concept of fundamental frequencies of bound orbits in Kerr spacetime introduced by Schmidt \cite{Schmidt02} and elaborated by Fujita and Hikida \cite{FujitaHikida09} based on a formulation of the geodesic equations in terms of the Mino time \cite{Mino03}. We also derived the post-Newtonian expansion of the effect which yields a more convenient formula and should still be sufficiently accurate for clocks moving in the gravitational field of the Earth. For the example of a GNSS-like satellite orbit compared to a geostationary orbit we found that the effect is of the order of $10^{-8}$ seconds per revolution and relative to the orbit time of order $10^{-12}$.  

The novel aspect of this generalized definition of the gravitomagnetic clock effect is that the two clocks may have arbitrary initial conditions and may follow completely different geodesics. This point is crucial if the effect should be tested with existing satellites or with a piggyback payload on another scientific mission. It also enables to consider the effect for astronomical objects. If, for example, two pulsars orbiting Sagittarius A* would be found, the gravitomagnetic clock effect could provide a consistency check of orbital data or of the value of the rotation parameter of the central black hole.

Here we considered geodesic motion in the Kerr spacetime, which is, of course, a very idealized situation. For a more realistic treatment it is certainly necessary to consider numerous perturbing effects, both of gravitational and nongravitational origin. Besides the stable clocks needed for a measurement of the gravitomagnetic clock effect a precise tracking of the clocks will be crucial.

\begin{acknowledgments}
We thank Hansjörg Dittus, Norman G\"urlebeck, Sven Herrmann, Bahram Mashhoon, and Volker Perlick for valuable discussions. Financial support from the German Research Foundation (DFG) through the research training group ``Models of Gravity'' is gratefully acknowledged. We also thank the Collaborative Research Center ``Relativistic geodesy and gravimetry with quantum sensors'' (geo-Q) for support.
\end{acknowledgments}

\appendix
\section{Calculation of fundamental frequencies}\label{app:FF}
The radial and latitudinal periods $\varLambda_r$ and $\varLambda_\theta$ as well as the fundamental frequencies $\Upsilon_\varphi$, $\Upsilon_t$, and $\Upsilon_\tau$ can be expressed in terms of Jacobian elliptic integrals. These are implemented in several computer algebra systems like Mathematica or Maple and can be calculated easily and quickly. In general, every integral of the form
\begin{align}
\int_a^b \frac{Q(z) dz}{\sqrt{P(z)}}\,,
\end{align}
where $Q$ is a rational function and $P$ is a polynomial of degree three or four, can be expressed in terms of elliptic integrals. If $P$ has only real zeros and $a$, $b$ are two neighbouring zeros of $P$, then a substitution of the form $z=\frac{\alpha n x^2+\beta}{nx^2+1}$ can be used to transform the above integral to the form
\begin{align}
C \int_0^1 \frac{Q(z(x)) dx}{\sqrt{(1-x^2)(1-k^2x^2)}}\,,
\end{align}
where $C$ is a constant and $0<k<1$. Now $Q(z(x))$ can be decomposed in partial fraction and the integral can be expressed in terms of complete elliptic integrals of the first, second, and third kind,
\begin{align}
K(k) & = \int_0^1 \frac{dx}{\sqrt{(1-x^2)(1-k^2x^2)}}\,,\nonumber\\
E(k) & = \int_0^1 \frac{(1-k^2x^2)dx}{\sqrt{(1-x^2)(1-k^2x^2)}}\,,\\
\Pi(n,k) & = \int_0^1 \frac{dx}{(1-nx^2)\sqrt{(1-x^2)(1-k^2x^2)}}\,.\nonumber
\end{align}

In our case we encounter the polynomials $R(\br)$ (see \eqref{dot r_lambda}) and with $\nu=\cos^2\theta$ in \eqref{dot theta_lambda}
\begin{align}
\left(\frac{d\nu}{d\blambda}\right)^2 & = 4 \ba^2 (1-E^2) \nu^3 + 4 (\bar K - (\ba E - \bar L_z)^2)\nu\\
& \quad + 4 (2\ba E(\ba E-\bar L_z) - \bar K - \epsilon \ba^2) \nu^2 =: \Theta_\nu(\nu)\,.\nonumber
\end{align}
For bound orbits $R$ has four real zeros $\br_{1}<\br_2<\br_{\rm p}<\br_{\rm a}$. All radial integrals can then be transformed to Jacobian elliptic integrals by the substitution $r=\frac{\alpha n x^2+\beta}{nx^2+1}$ with $\alpha=\br_2$, $\beta=\br_{\rm p}$, and $n=-\frac{\br_{\rm a}-\br_{\rm p}}{\br_{\rm a}-\br_2}$. The radial period is then given by
\begin{align}
\varLambda_r & = \frac{4K(k_r)}{\sqrt{(1-E^2)(\br_{\rm p}-\br_1)(\br_{\rm a}-\br_2)}}\,,\\
k_r^2 & = \frac{(\br_{\rm a}-\br_{\rm p})(\br_2-\br_1)}{(\br_{\rm a}-\br_2)(\br_{\rm p}-\br_1)}\,.\nonumber
\end{align}
 
For bound orbits $\Theta_\nu$ has three real zeros $0=\nu_0<\nu_{\rm max}<1<\nu_1$. With $\nu=\frac{\alpha n x^2+\beta}{nx^2+1}$ where $\beta=\nu_{\rm max}$, $\alpha=\nu_1$, and $n=-\frac{\nu_{\rm max}}{\nu_1}$ we find
\begin{align}
\varLambda_\theta & = \frac{4K(k_\theta)}{\sqrt{\ba^2(1-E^2)\nu_1}}\,, \quad k_\theta^2=\frac{\nu_{\rm max}}{\nu_1}\,.
\end{align}

In the same way we may transform the integrals appearing in the definitions of $\Upsilon_\varphi$ \eqref{Upsilonphi}, $\Upsilon_t$ \eqref{Upsilont}, and $\Upsilon_\tau$ \eqref{Upsilontau}. We find
\begin{align}
\Upsilon_\varphi & = \frac{1}{K(k_r)} \int_0^1 \frac{\Phi_r(x)dx}{\sqrt{(1-x^2)(1-k_r^2x^2)}}\nonumber\\
&\quad  + \frac{1}{K(k_\theta)} \int_0^1 \frac{\Phi_\theta(x)dx}{\sqrt{(1-x^2)(1-k_r^2x^2)}}\,,
\end{align}
where
\begin{align}
\Phi_r(x) & = \frac{\ba(\br_{\rm p}-\br_2)}{(h_1-h_2)} \sum_{i=1}^2 \frac{(\br_{\rm p}-h_i)^{-1}(-1)^{i}\mathcal{R}(h_i)}{(\br_2-h_i)(1-N_{r,i}x^2)}\nonumber\\
& \quad + \frac{\ba\mathcal{R}(\br_2)}{\Delta_{\br_2}}\,, \quad N_{r,i} = \frac{(\br_{\rm a}-\br_{\rm p})(\br_2-h_i)}{(\br_{\rm a}-\br_2)(\br_{\rm p}-h_i)}\,,
\end{align}
with the horizons $h_{1,2}=1\pm \sqrt{1-\ba^2}$ and
\begin{align}
\Phi_\theta(x) & = \frac{\mathcal{T}(\nu_1)}{\nu_1-1} + \frac{\bL(\nu_1-\nu_{\rm max})}{(\nu_1-1)(1-\nu_{\rm max})(1-N_\theta x^2)}\,,\nonumber\\
&\quad N_\theta=\frac{\nu_{\rm max}(1-\nu_1)}{\nu_1(1-\nu_{\rm max})}\,.
\end{align}
In terms of Jacobian elliptic integrals this reads
\begin{widetext}
\begin{align}
\Upsilon_\varphi & = \frac{\ba(\br_{\rm p}-\br_2)}{(h_1-h_2)} \sum_{i=1}^2 \frac{(-1)^{i}\mathcal{R}(h_i)}{(\br_2-h_i)(\br_{\rm p}-h_i)} \frac{\Pi(N_{r,i},k_r)}{K(k_r)} + \frac{\ba\mathcal{R}(\br_2)}{\Delta_{\br_2}} + \frac{\mathcal{T}(\nu_1)}{\nu_1-1} + \frac{\bL(\nu_1-\nu_{\rm max})}{(\nu_1-1)(1-\nu_{\rm max})} \frac{\Pi(N_\theta,k_\theta)}{K(k_\theta)}\,. \label{ellYphi}
\end{align}
Analogously we get
\begin{align}
\Upsilon_\tau & = \ba^2 \left[ \nu_1 - \frac{(\nu_1-\nu_{\rm max})}{1-k_\theta^2} \frac{E(k_\theta)}{K(k_\theta)} \right] + \br_2^2+2\br_2(\br_p-\br_2) \frac{\Pi(N_{r,3},k_r)}{K(k_r)} \nonumber\\
& \quad - \frac{(\br_p-\br_2)^2}{2(1-N_{r,3})} \bigg[1- \frac{N_{r,3}E(k_r)}{(N_{r,3}-k_r^2)K(k_r)}+ \frac{(N_{r,3}^2+3k_r^2-2N_{r,3}-2N_{r,3}k_r^2)\Pi(N_{r,3},k_r)}{(N_{r,3}-k_r^2)K(k_r)} \bigg] \label{ellYtau}
\end{align}
and
\begin{align}
\Upsilon_t & = \ba \mathcal{T}(\nu_1) - \frac{\ba^2 E (\nu_1-\nu_{\rm max})}{1-k_\theta^2} \frac{E(k_\theta)}{K(k_\theta)} + \frac{(\br_2^2+\ba^2)\mathcal{R}(\br_2)}{\Delta_{\br_2}} + \frac{\ba(\br_{\rm p}-\br_2)}{(h_1-h_2)} \sum_{i=1}^2 \frac{(-1)^{i}(h_i^2+\ba^2)\mathcal{R}(h_i)}{(\br_2-h_i)(\br_{\rm p}-h_i)} \frac{\Pi(N_{r,i},k_r)}{K(k_r)}\nonumber\\
& \quad - \frac{E(\br_{\rm p}-\br_2)^2}{2(1-N_{r,3})} \bigg[1- \frac{N_{r,3}E(k_r)}{(N_{r,3}-k_r^2)K(k_r)}+ \frac{(N_{r,3}^2+3k_r^2-2N_{r,3}-2N_{r,3}k_r^2)\Pi(N_{r,3},k_r)}{(N_{r,3}-k_r^2)K(k_r)} \bigg]\nonumber\\
& \quad +2E(\br_{\rm p}-2)(\br_2+1) \frac{\Pi(N_{r,3},k_r)}{K(k_r)}\,, \label{ellYt}
\end{align}
\end{widetext}
where
\begin{align}
N_{r,3} & = \frac{\br_{\rm a}-\br_{\rm p}}{\br_{\rm a}-\br_2}\,.
\end{align}

\section{Details of the post-Newtonian expansion}\label{app:PN}

To determine the post-Newtonian expansion of the general gravitomagnetic clock effect \eqref{deltatau} we first need an expansion for small $\ba$ of the constants of motion $E$, $\bL_z$, and $\bK$ as well as the zeros $\br_1$, $\br_2$, and $\nu_1$. To find this expansion we consider them as functions of $\ba$ and compare the coefficients in 
\begin{align}
0 & = R(\br)-(1-E^2)(\br_{\rm a}-\br)(\br-\br_{\rm p})(\br-\br_2)(\br-\br_1)\,,\\
0 & = \Theta_\nu(\nu) - 4\ba^2(1-E^2)\nu(\nu_{\rm max}-\nu)(\nu_1-\nu)\,,
\end{align}
taking into account that $\nu_1=c_2\ba^{-2}+c_1\ba^{-1}+\mathcal{O}(\ba^0)$ for some constants $c_2$, $c_1$. Without loss of generality $E$ can be assumed as positive and we find
\begin{align}
E & \approx \sqrt{\frac{4(1-e^2)+\bp(\bp-4)}{\bp(\bp-3-e^2)}} \mp \ba \frac{(1-e^2)^2\cos i}{\bp(\bp-3-e^2)^\frac32}\,,\\
\bL_z & \approx \pm \frac{\bp\cos i}{\sqrt{\bp-3-e^2}} \mp \ba \frac{(3+e^2)\cos^2i}{\bp^\frac12 (\bp-3-e^2)^\frac32}\times\nonumber\\
& \quad \times \sqrt{(\bp-2)^2-4e^2}\,,
\end{align}
\begin{align}
\bK & \approx \frac{\bp^2}{\bp-3-e^2} \pm \ba \frac{2\bp^\frac32\cos i\sqrt{(\bp-2)^2-4e^2}}{(\bp-3-e^2)^2}\,,\\
\br_2 & \approx \frac{2\bp}{\bp-4} \mp \ba \frac{4\cos i\sqrt{\bp((\bp-2)^2-4e^2)}}{(\bp-4)^2}\,,\\
\br_1 & = \mathcal{O}(\ba^2)\,,\\
\nu_1 & \approx \frac{\bp^3}{\ba^2(\bp-4)(1-e^2)} \mp \frac{8\bp^\frac32\cos i}{\ba(1-e^2)(\bp-4)^2} \times \nonumber\\
& \quad \times \sqrt{(\bp-2)^2-4e^2}\,,
\end{align}

where $\br_{\rm p}=\bp(1+e)^{-1}$, $\br_{\rm a}=\bp(1-e)^{-1}$, and $\nu_{\rm max}=\sin^2i$. Here the upper sign corresponds to prograde and the lower sign to retrograde motion. 

These expansions must then be inserted in the expressions \eqref{ellYphi} and \eqref{ellYtau} to derive the post-Schwarzschild expansion of the gravitomagnetic clock effect. To first order in $\ba$ we get
\begin{widetext}
\begin{align}
\Upsilon_{\varphi} & \approx \pm \frac{\bp}{\sqrt{\bp-3-e^2}} + \ba \frac{\sqrt{(\bp-2)^2-4e^2}}{\sqrt{\bp-3-e^2}} \bigg[ \frac{\bp^\frac12}{4}-\frac{\bp^\frac12(\bp-6-2e)\Pi(n_1,k)}{4(\bp-2-2e)K(k)} - \frac{\bp-3-e^2+\cos i(3+e^2)}{\bp^\frac12(\bp-3-e^2)} \bigg]\,, \label{exp_Yphi}\\
\Upsilon_\tau & \approx \frac{\bp^2}{(\bp-4)(1-e^2)} \bigg[ \frac{12\bp-4e^2-\bp^2-28}{2(\bp-4)} + \frac{(\bp-6+2e)E(k)}{2K(k)} + \frac{(\bp-6-2e)(\bp-3-e^2)\Pi(n_3,k))}{(\bp-4)(1+e)K(k)} \bigg] \nonumber\\
& \quad \pm \ba \frac{\bp^\frac32\cos i\sqrt{(\bp-2)^2-4e^2}}{(\bp-4)(1-e^2)} \bigg[ \frac{12e^2+\bp^2-12\bp+20}{2(\bp-4)^2} - \frac{(\bp-6+2e)E(k)}{(\bp-4)K(k)} + \frac{(\bp-6+2e)E^2(k)}{2(\bp-6-2e)K^2(k)} \nonumber\\
& \quad - \frac{(\bp-6-2e)(\bp-3e^2-1)\Pi(n_3,k))}{(\bp-4)^2(1+e)K(k)} + \frac{(\bp-3-e^2)\Pi(n_3,k)E(k)}{(\bp-4)(1+e)K^2(k)} \bigg]\,, \label{exp_Ytau}
\end{align}
\end{widetext}
where $\cos i = \sqrt{1-\nu_{\rm max}} = \sin\theta_{\rm max}$ and
\begin{align}
k^2 & =k_r^2(\ba=0)=\frac{2e}{\bp-6+2e}\,,\\
n_1 & =N_{r,1}(\ba=0)=\frac{16e}{(\bp-2e-2)(\bp-6+2e)}\,,\\
n_3 & =N_{r,3}(\ba=0)=\frac{2e(\bp-4)}{(1+e)(\bp-6+2e)}\,.
\end{align}
Note that the expansion \eqref{exp_Yphi} is not valid for polar orbits. For this special case, it is $L_z=0$, and we find from the definition \eqref{defYphi},
\begin{align}
\Upsilon_\varphi = \frac{2}{\varLambda_r} \int_{\br_{\rm p}}^{\br_{\rm a}} \frac{\Phi_r(\br)d\br}{\sqrt{R(\br)}} - \ba E\,.
\end{align}
For $\ba=0$, we, therefore, get for polar orbits $\Upsilon_\varphi=0$, as expected. It is, therefore, not possible to interchange the limits $\nu_{\rm max}\to1$ and $\ba\to0$ in $\Upsilon_\varphi$.

Subsequently, we insert $\bp=p/M$ in \eqref{exp_Yphi} and \eqref{exp_Ytau} and consider the limit $M\to 0$ to derive the post-Newtonian expansion \eqref{taugm} of $\tau(\pm2\pi)$.

\bibliographystyle{unsrt}
\bibliography{clockeffect}

\end{document}